\documentclass[applsci,article,accept,moreauthors,pdftex]{Definitions/mdpi} 
\firstpage{1} 
\pubvolume{xx}
\issuenum{1}
\articlenumber{5}
\pubyear{2019}
\copyrightyear{2019}
\history{Received: 26 September 2019 ; Accepted: 2 October 2019 ; Published: date}
\Title{Topological Phase Diagram of BiTeX--Graphene Hybrid Structures}

\Author{Zoltán Tajkov %Please carefully check the accuracy of names and affiliations.
$^{1,}$*\orcidA{}, Dávid Visontai $^{2}$\orcidB{}, László Oroszlány $^{3,\dagger}$\orcidC{} and János Koltai $^{1}$\orcidD{}}

\AuthorNames{Zoltán Tajkov, Dávid Visontai, László Oroszlány and János Koltai}

\address{%
$^{1}$ \quad Department of Biological Physics, ELTE E{\" o}tv{\" o}s Loránd University, P{\'a}zmány P{\'e}ter s{\' e}t{\' a}ny 1/A, 1117~Budapest,~Hungary;  koltai@angel.elte.hu\\
$^{2}$ \quad Department of  Material Physics, ELTE E{\" o}tv{\" o}s Loránd University, P{\'a}zmány P{\'e}ter s{\' e}t{\' a}ny 1/A, 1117~Budapest,~Hungary; david.visontai@complex.elte.hu\\
$^{3}$ \quad Department of Physics of Complex Systems, ELTE E{\" o}tv{\" o}s Loránd University, P{\'a}zmány P{\'e}ter s{\' e}t{\' a}ny 1/A, 1117 Budapest, Hungary; oroszl@elte.hu
}

\corres{Correspondence: novidad21@caesar.elte.hu}
\firstnote{MTA-BME Lendület Topology and Correlation Research Group, Budafoki {\'u}t 8., H-1111
Budapest, Hungary.}
\abstract{Combining graphene with other novel layered materials is a possible way for
engineering the band structure of charge carriers. Strong spin-orbit coupling in BiTeX
compounds and the recent fabrication of a single layer of BiTeI points towards a feasible
experimental realization of a Kane--Mele phase in graphene-based heterostructures.
Here, we theoretically demonstrate the tunability of the topological phase of hybrid
systems built from graphene and BiTeX (X = I, Br, Cl) layers by uniaxial in-plane tensile and out-of plane compressive strain.
We show that structural stress inherently present in fabricated samples could induce a topological phase
transition, thus turning the sample in a novel experimental realization of a time reversal
invariant topological insulator.}


\begin{document}
%%%%%%%%%%%%%%%%%%%%%%%%%%%%%%%%%%%%%%%%%%

%%%%%%%%%%%%%%%%%%%%%%%%%%%%%%%%%%%%%%%%%%

\section{Introduction}
After the first successful isolation of graphene in 2004, graphene has proved to be an excellent template material with outstanding mechanical properties for revolutionary
applications \cite{novoselov_electric_2004}. Due to the small atomic weight of the carbon atoms building up the graphene lattice, the effect of spin-orbit coupling (SOC) on the electronic band structure is negligible \cite{castro_neto_impurity-induced_2009}. This makes graphene a semimetal. There has recently been a strong push to find ways to enhance spin-orbit coupling in graphene \cite{han_graphene_2014} in order to enable spintronics
applications \cite{zutic_spintronics:_2004}. Theoretical investigations showed that one can overcome this limitation in several ways. Introducing curvature in the graphene sheet {enhances} the effect of the spin-orbit interaction up to $\approx 17\, \mathrm{meV}$ in realistic situations \cite{curved_graphene}. It has been also shown that impurities influence the SOC in a positive way and can amplify its strength up to $\approx 7\, \mathrm{meV}$ \cite{castro_neto_impurity-induced_2009}. 
A more practical approach, from an engineering point of view, is to use 2D heterostructures in order to introduce spin-orbit interaction into graphene, since the already existing and understood fabricating procedures \cite{geim_van_2013} and  van der Waals heterostructures provide more
robust control towards technological reproducibility of devices.
SOC can be {introduced to} graphene by van der Waals constituents, which itself is characterized by strong instrinsic spin-oribt interaction. Prospective candidates are bismuth tellurohalides (e.g., BiTeX with X = I, Br, and Cl). 
The key element of these compounds is bismuth with strong atomic SOC. The structure element of BiTeX is a trilayer with the X-Bi-Te stacking. The already large intrinsic SOC in Bi
and the structural asymmetry combined with a large in-plane gradient of the crystal field
in this lattice results in giant Rashba spin-splitting semiconductors \cite{ast_giant_2007,wu_enhanced_2015,sakano_three-dimensional_2012,butler_mapping_2014,kong_polarity_2018}. 
A single trilayer of BiTeI, that shows the Rashba spin-split band structure can be exfoliated from the bulk BiTeI as it was recently shown by Fülöp {et al.}, using a novel exfoliation technique \cite{fulop_exfoliation_2018}. It was also shown theoretically that centrosymmetric thin films
composed from topologically trivial BiTeI trilayers are quantum spin Hall insulators and
properly stacked bulk compounds of BiTeX results in topological insulating phase \cite{eremeev_new_2015,nechaev_quantum_2017}. In monolayer BiTeI the band gap is increased, compared to the bulk material, still the basic characteristics are preserved, hence it is a good candidate as a SOC inducing component in graphene-based heterostructures, as it was previously proposed by earlier works \cite{tajkov_transport_2017,kou_robust_2014}.

The fact that fabricated devices are usually subject to mechanical stress requires the investigation of the impact of such effects on the electronic properties of the systems. In experimental setups a considerable effort has been made to control and manipulate the electronic and optical properties of novel heterostructures by strain fields \cite{CastellanosGomez_Strain_Engineering_MoS2_2013,Jiang_Pseudomagnetic,Sanchez_strain_tuning_2017,Goldsche_Strain_Fields_graphene_2018}. 

To realize mechanically driven 2D $\mathbb{Z}_2$ topological insulators, several theoretical proposals have been made. For example, in two-dimensional dumbbell stanene a topological phase can be tuned by compression and in hydrogenated ultra-thin tin films a topological gap can be opened by mechanical distortions. \cite{ren_topological_2016}

{As we claimed in our previous work, the single sided heterostructures of BiTeX and graphene under appropriate conditions are topological insulators, however we showed evidence for only the BiTeBr--graphene structure \cite{tajkov_uniaxial_2019}. In this manuscript we investigate in details the phase diagram of the BiTeCl--graphene and BiTeI--graphene systems in the level of density functional theory calculations, as well as the BiTeBr--graphene structure. We analyze and discuss the differences of the electronic properties of the heterostructures.}

\section{Results}

\begin{figure}[H]
\centering
\includegraphics[width=10cm]{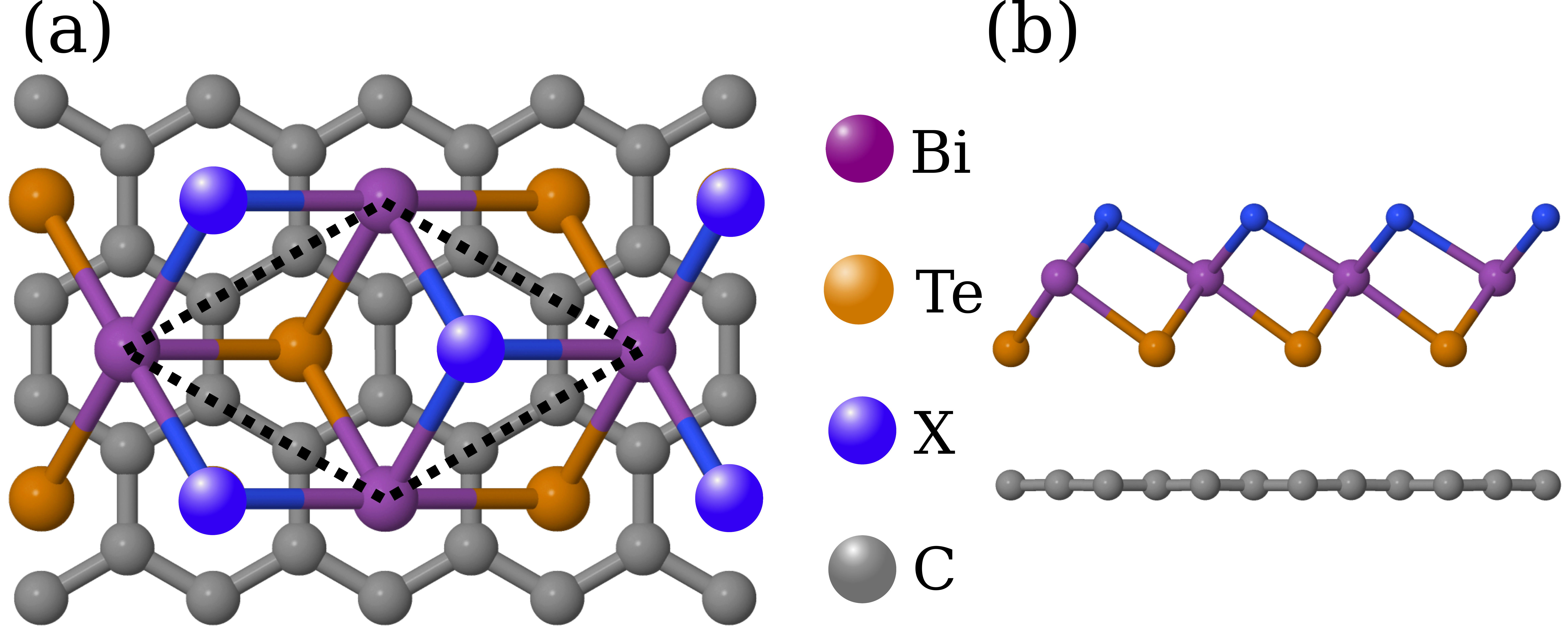}
\caption{Top (\textbf{a}) and side (\textbf{b}) view of the BiTeX--graphene (X = I, Br, Cl)
heterostructure. The black dashed line indicates the unit cell. (color online).} 
\label{structure}
\end{figure}

To set up the investigated systems, the experimental characteristics of the BiTeX crystals were used. The in-plane lattice
constants are $a_{\mathrm{BiTeX}}$ = 4.34 \AA, 4.24 \AA, 4.27 \AA $\ $ for X = I, Br, Cl respectively \cite{shevelkov_crystal_1995}.
Inasmuch as these values are close to a $\sqrt{3}\times a_{\mathrm{Gr}}$, a $30^{\circ}$ rotated, $\sqrt{3}\times \sqrt{3}$ unit cell of graphene was placed on top of the BiTeX layer as a starting point for the calculations.
This choice of configuration carries a stress between the two systems due to the lattice mismatch. The scale of discrepancy is $a_{\mathrm{BiTeX}} / a_{\sqrt{3}\times\sqrt{3}\mathrm{Gr}} = 1.86 \%, -0.49\%, 0.21\%$ for X = I, Br, Cl respectively. Performing a full geometry optimization on the hybrid systems it shows that the crystal structure of graphene is preserved and the lattice constant of BiTeX layer was adjusted.
 This phenomenon is comprehensible considering the fact that the bonds in the BiTeX crystals have more degrees of freedom to accommodate. Hence, this reasonable strain only effects the BiTeX in the heterostructures, as it {may alter} the band structure of the BiTeX layers, but it does not influence our main conclusions as we focus on the Dirac cones of the graphene sheet.
The supercell is depicted in Figure \ref{structure}, consisting of six carbon atoms, one bismuth, one tellurium, and one halogen atom. As it was already shown previously, the most stable horizontal configuration is the so called hollow, when the adjacent atom of the BiTeX layer ({i.e.}, Te or X) is positioned above the center of a hexagon of carbon atoms in the graphene layer \cite{fulop_exfoliation_2018,kou_robust_2014}. The structures in I-Bi-Te--C configurations are characterized by lower total energy than those in the structural model with the Te-Bi-I--C sequence \cite{kou_robust_2014}.

\begin{figure}[H]
\centering
    \includegraphics[width=\textwidth]{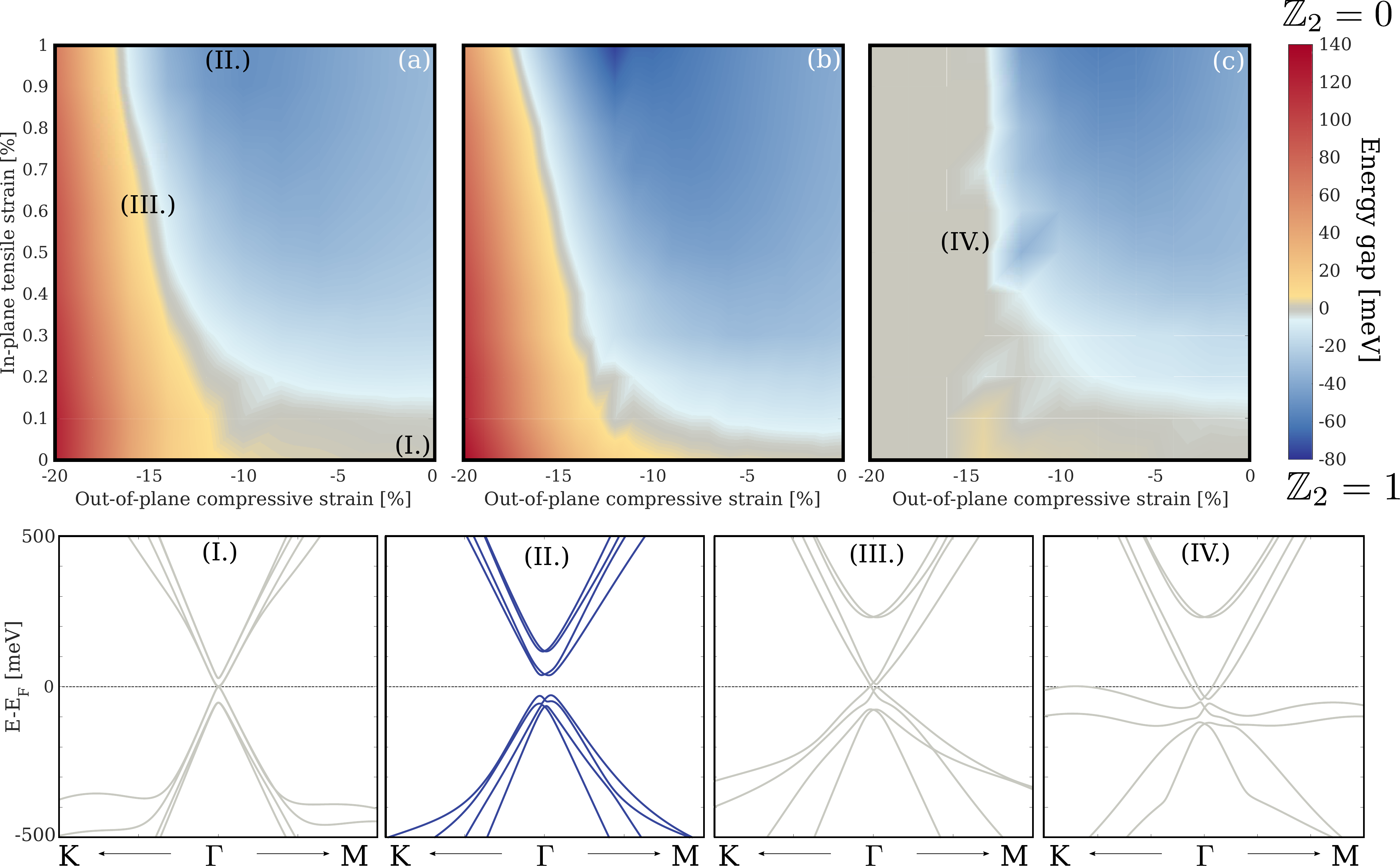}
\caption{The calculated in-plane tensile--out-of-plane compressive strain phase diagram of the BiTeCl/BiTeBr/BiTeI--graphene structures (\textbf{a}--\textbf{c}) respectively. The blue colors indicate that the gap is topological, while the red hue means that is trivial. The metallic shade marks the phase boundary. Subfigures (I.)--(IV.) are the corresponding band structures of the marked phases near the $\Gamma$ point in momentum space. (color online).} 
\label{phase}
\end{figure}

In order to simulate the effect of tensile uniaxial and uniform strain we stretched the in-plane unit cell vectors parallel to a carbon-carbon bond and let the atomic positions in the constrained unit cell relax. The compressive strain was taken in consideration by reducing the distance between the BiTeX and the graphene layers. In these calculations we did not {let the atomic positions relax}. As we applied both types of strain to the system we calculated the size of the band gap and the topological index in every case. In Figure \ref{phase} we show the corresponding phase diagrams. In relevant cases we include the band structures as well. The sign of the band gap indicates the topological invariant, it is negative if the system is topological and positive otherwise.

The first subfigure (a) presents the calculations on the BiTeCl--graphene heterostructure. At the optimal geometry the system is semi-metallic due to the touching altered Dirac cones of graphene as it can be seen in subfigure (I.). As the tensile strain increases, a topologically non-trivial band gap opens in graphene (subfigure (II.)). Turning on the compressive out-of-plane strain, it widens the band gap in the beginning, but later closes it (see the corresponding band structure in subfigure (III.)) and changes its topological favor to trivial. In this scenario, a $50 \, \mathrm{meV}$ topological and a $120 \, \mathrm{meV}$ trivial gap can be reached. The second subfigure (b) corresponds to the BiTeBr--graphene system. The same statements can be made with a slightly different energy scale. {The behavior of the band gap is similar to the case of the BiTeCl--graphene structure. At fixed tensile strain, applying pressure widens the topological gap in the beginning up to $70 \, \mathrm{meV}$, but later closes the gap and the reopened gap is trivial, which can be as high as $140 \, \mathrm{meV}$}.  The third subfigure (c) shows the BiTeI--graphene structure. The effect of in-plane strain is the same as above: it promotes the topological phase. The consequence of the compressive strain is somewhat different. This system turns metallic instead of a trivial insulator due to some non graphene bands reaching the Fermi level as pressure is increased (see subfigure (IV.)). 

{The deviating behavior of the BiTeI--graphene structure can be explained by the different work function of the BiTeX materials. In the case of the Te terminated BiTeBr and BiTeCl, the work function is around 4.7--4.5 {eV} as it was calculcated by Fiedler {et al.}  \cite{fiedler_termination_2015}. This is close to the theoretically calculated value for graphene (4.6 {eV} Yu {et al.} \cite{yu_tuning_2009}). The work function of Te terminated BiTeI is larger by almost 0.5~{eV}. The altered Dirac cones of graphene can be distinguished in the band structure of the hybrid systems because the cones are placed in the band gap of the BiTeX systems due to the similar work functions. As we stated in our previous work, the hybrid structure consisting graphene and X terminated BiTeX does not provide this nature \cite{tajkov_uniaxial_2019}, which is in agreement with the significantly larger work function of X terminated BiTeCl and BiTeBr (6.2--6.0 {eV} \cite{fiedler_termination_2015}). In the case of the BiTeI--graphene structure the altered Dirac cones are placed closer to the occupied bands of BiTeI due to the larger value of the work function. As we apply pressure those bands populate the band gap at a higher rate as in the case of BiTeBr or BiTeCl, resulting in a metallic phase.}

Based on the presented results we conclude that both type of mechanical distortions have
a striking effect on the band gap, however they favor different topological phases. Out-of-plane strain opens a trivial band gap, while in-plane strain drives the
system into the topological phase.

We would like to point out that the largest compressive strain we applied during our calculations would correspond to a nominal pressure of 20 GPa. We estimated this value as the derivation of the total energy per unit area over the reduced distance. In modern experimental setups these mechanical stresses can be routinely achieved \cite{kleppe_new_2014,ni_uniaxial_2008,braganza_hydrostatic_2002,vos_high-pressure_1992,kullmann_effect_1984}.

%%%%%%%%%%%%%%%%%%%%%%%%%%%%%%%%%%%%%%%%%%
\section{Methods}

The optimized geometry and ground state Hamiltonian and overlap matrix elements of each
structure were self consistently obtained by the SIESTA implementation of density functional
theory (DFT) \cite{artacho_siesta_2008,noauthor_siesta_2002}. {For every calculation, the spin-orbit interaction was included as it was implemented in SIESTA by Dr. Ramón Cuadrado based on the original on-site SO formalism and implementation developed by Prof. Jaime Ferrer \cite{fernandez-seivane_site_2006}}. SIESTA employs norm-conserving pseudopotentials to account for the
core electrons and linear combination of atomic orbitals to construct the valence states. We used the pseudopotentials optimized by Rivero {et al.} \cite{rivero_systematic_2015}. For all cases, the considered samples were separated with a minimum of 18.5 \AA$\, $ thick vacuum in the perpendicular direction. The generalized gradient approximation of the exchange
and the correlation functional was used with Perdew--Burke--Ernzerhof parametrization \cite{perdew_generalized_1996}, {as the pseudopotentials were created to the PBE functionals}, with a double-$\zeta$ polarized basis set
and a real-space grid defined with an equivalent energy cutoff of 1000 Ry. The Brillouin
zone integration was sampled by a 24 $\times$ 24 $\times$ 1 Monkhorst--Pack $\mathbf{k}$-grid \cite{monkhorst_special_1976}. The geometry
optimizations were performed until the forces were smaller than 0.01 eV/\AA. 

The choice of
pseudopotentials optimized by Rivero {et al.} ensures that both the obtained geometrical
structures and the electronic band properties are reliable. As a benchmark we validated our
method by comparing the electronic properties of the bulk BiTeI with the experimental data.
This approach gave us 130 meV band gap and 4.6 eV\AA$\,$ as Rashba parameter for BiTeI bulk.
The corresponding experimental results are 190 meV and 3.8 eV\AA$\,$ respectively \cite{ishizaka_giant_2011}.

We extracted the topological phase information of the systems by determining the Wannier center flow. The center of the Wannier function can be expressed as the phase of eigenvalues of a matrix obtained as the product of the Berry connection along the ``Wilson loop''  {by our own post-processing tool}.  { The used real space Hamiltonian was calculated by the SIESTA self-consistent cycle.} The $\mathbb{Z}_2$ topological numbers  {were} expressed as the number of times mod 2 of the partner switching of these phases during a complete period of the ``time reversal pumping'' process \cite{asboth_short_2016,fu_topological_2011,yu_equivalent_2011}.

%%%%%%%%%%%%%%%%%%%%%%%%%%%%%%%%%%%%%%%%%%
\section{Conclusions}

In summary, we have explored the intriguing topological phase diagrams of bismuth tellurohalide/graphene
heterostructures by means of first principles calculations. We showed that the in-plane uniaxial tensile strain opens a topologically non-trivial band gap in graphene in the vicinity of a material features strong inherent spin-orbit coupling. On the other hand, the compressive stress on the device promotes a trivial band gap. {The BiTeBr--graphene and BiTeCl--graphene systems can be tuned from topological to trivial state and vica versa by applying mechanical stress. Furthermore, the BiTeI--graphene structure can be tuned from topological state to metallic. These systems lead to a novel realization of the time reversal invariant topological insulating phase, thus making these heterostructures potential candidates for quantum technology applications without the necessity of low temperature.}

%%%%%%%%%%%%%%%%%%%%%%%%%%%%%%%%%%%%%%%%%%

%%%%%%%%%%%%%%%%%%%%%%%%%%%%%%%%%%%%%%%%%%
\vspace{6pt} 

%%%%%%%%%%%%%%%%%%%%%%%%%%%%%%%%%%%%%%%%%%
%% optional
%\supplementary{The following are available online at \linksupplementary{s1}, Figure S1: title, Table S1: title, Video S1: title.}


%%%%%%%%%%%%%%%%%%%%%%%%%%%%%%%%%%%%%%%%%%
\authorcontributions{Conceptualization, J.K. and L.O.; methodology, Z.T.; software, Z.T. and D.V.; validation, D.V.; writing--original draft preparation, Z.T.; writing--review and editing, J.K., L.O. and Z.T.; supervision, J.K. and L.O.}

%%%%%%%%%%%%%%%%%%%%%%%%%%%%%%%%%%%%%%%%%%
\funding{This research was supported by the National Research, Development and Innovation Fund of Hungary within the Quantum Technology National Excellence Program (Project Nr. 2017-1.2.1-NKP-2017-00001); grants no. K112918, K115608, FK124723 and K115575. This work was completed in the ELTE Excellence Program (1783-3/2018/FEKUTSRAT) supported by the Hungarian Ministry of Human Capacities. JK acknowledge the Bolyai and Bolyai+ program of the Hungarian Academy of Sciences.} 

%%%%%%%%%%%%%%%%%%%%%%%%%%%%%%%%%%%%%%%%%%
\acknowledgments{We acknowledge [NIIF] for awarding us access to resource based in
Hungary at Debrecen.}

%%%%%%%%%%%%%%%%%%%%%%%%%%%%%%%%%%%%%%%%%%
\conflictsofinterest{The authors declare no conflict of interest.}

%%%%%%%%%%%%%%%%%%%%%%%%%%%%%%%%%%%%%%%%%%
%% optional

%%%%%%%%%%%%%%%%%%%%%%%%%%%%%%%%%%%%%%%%%%
\bibliographystyle{mdpi}
\reftitle{References}

%=====================================
% References, variant B: internal bibliography
%=====================================

% The following MDPI journals use author-date citation: Arts, Econometrics, Economies, Genealogy, Humanities, IJFS, JRFM, Laws, Religions, Risks, Social Sciences. For those journals, please follow the formatting guidelines on http://www.mdpi.com/authors/references
% To cite two works by the same author: \citeauthor{ref-journal-1a} (\citeyear{ref-journal-1a}, \citeyear{ref-journal-1b}). This produces: Whittaker (1967, 1975)
% To cite two works by the same author with specific pages: \citeauthor{ref-journal-3a} (\citeyear{ref-journal-3a}, p. 328; \citeyear{ref-journal-3b}, p.475). This produces: Wong (1999, p. 328; 2000, p. 475)

%%%%%%%%%%%%%%%%%%%%%%%%%%%%%%%%%%%%%%%%%%
%% optional

%% for journal Sci
%\reviewreports{\\
%Reviewer 1 comments and authors’ response\\
%Reviewer 2 comments and authors’ response\\
%Reviewer 3 comments and authors’ response
%}

%%%%%%%%%%%%%%%%%%%%%%%%%%%%%%%%%%%%%%%%%%
\end{document}